\documentclass{PoS}
\usepackage{booktabs}
\usepackage{amsmath}
\usepackage{physics}
\usepackage{marginnote}
\usepackage{simplewick}
\usepackage{tikz}
\usepackage{graphics}
\usepackage{gincltex}
\usepackage{bm}
\usepackage[backend=biber,bibencoding=utf8,url=false,doi=false,eprint=false,giveninits=true,style=numeric,sorting=none,maxnames=2]{biblatex}
\renewbibmacro{in:}{}
\AtEveryBibitem{\clearfield{title}}
\newcommand{\be}{\begin{equation}}
\newcommand{\ee}{\end{equation}}
\newcommand{\bea}{\begin{eqnarray}}
\newcommand{\eea}{\end{eqnarray}}
\newcommand{\bes}{\begin{eqnarray}}
\newcommand{\ees}{\end{eqnarray}}
\newcommand{\ba}{\begin{array}}
\newcommand{\ea}{\end{array}}

\newcommand{\id}{I}

\renewcommand{\i}{\mathrm{i}}

\title{Frequency-splitting estimators for single-propagator traces}

\ShortTitle{Frequency-splitting estimators}

\author{Leonardo~Giusti$^a$, \speaker{Tim~Harris}$^a$, {Alessandro~Nada}$^b$, Stefan~Schaefer$^b$\\
        \llap{$^a$}Dipartimento di Fisica, Universit\`a di Milano-Bicocca, and INFN, Sezione di Milano-Bicocca\\
        Piazza della Scienza 3, I-20126 Milano, Italy\\
        \llap{$^b$}John von Neumann Institute for Computing, DESY\\
        Platanenallee 6, D-15738 Zeuthen, Germany\\
        E-mail: \email{leonardo.giusti@mib.infn.it}, \email{tim.harris@mib.infn.it},\email{alessandro.nada@desy.de}, \email{stefan.schaefer@desy.de}}

\abstract{
    In these proceedings we address the computation of quark-line disconnected
    diagrams in lattice QCD.
    The evaluation of these diagrams is required for many phenomenologically
    interesting observables, but suffers from large statistical errors due to
    the vacuum and random-noise contributions to their variances.
    Motivated by a theoretical analysis of the variances, we introduce
    a new family of stochastic estimators of single-propagator traces built
    upon a frequency splitting combined with a hopping expansion of the quark
    propagator, and test their efficiency in two-flavour QCD with pions as
    light as 190 MeV.
    The use of these estimators reduces the cost of the computation by one
    to two orders of magnitude over standard estimators depending on the
    fermion bilinear.
    As a concrete application, we show the impact of these findings on the
    computation of the hadronic vacuum polarization contribution to the muon
    anomalous magnetic moment.
    \vskip 1cm
    DESY 20-010
}

\FullConference{37th International Symposium on Lattice Field Theory - Lattice2019\\
		16-22 June 2019\\
		Wuhan, China}

\begin{document}

\section{Introduction}

Quark-line disconnected Wick contractions are ubiquitous in lattice QCD, and
those involving single-propagator traces contribute to many interesting
observables, such as Standard Model processes like $K\rightarrow\pi\pi$,
strong and electromagnetic isospin-breaking corrections and isosinglet
spectroscopy.
These observables are particularly computationally challenging as they can
have large vacuum contributions to their variance, as well as large random
noise contributions from the auxiliary fields introduced to evaluate them
stochastically~\cite{Bitar:1988bb,Michael:1998sg}.
The former can be ameliorated by multi-level
integration~\cite{Ce:2016idq,Ce:2016ajy}, whereas in the following, we
suppress the latter contribution by introducing a new family of
reduced-variance stochastic estimators which are constructed using a
frequency splitting and a hopping expansion applied at a large quark mass.
These estimators can be applied in standard Monte Carlo simulations, and here
we report on numerical tests with $N_\mathrm{f}=2$ O($a$)-improved Wilson
fermions, where we observe between one and two orders of magntitude reduction
in the variance (or cost), depending on the fermion bilinear considered.
Further details and any unexplained notation can be found in
ref.~\cite{Giusti:2019kff}.

\section{Variances of single-propagator traces}
\label{sec:var}
It is often sufficient to study the variance of individual disconnected Wick
contractions, e.g. $W_1$ and $W_0$ (assumed to be real), whose product
comprises a larger correlation function
\begin{align}
    {\cal C}_{_{W_1 W_0}}(x_1,x_0) &=  \Big\langle \Big[W_1(x_1) - \langle
    W_1(x_1) \rangle \Big]
    \Big[W_0(x_0) - \langle W_0(x_0) \rangle \Big] \Big\rangle\; . 
\end{align}
For large $\lvert x_1-x_0\rvert$ the variance factorizes into the product of the
variances of the individual contractions,
\begin{align}
    \sigma^2_{_{{\cal C}_{W_1 W_0}}}(x_1,x_0) &\approx
    \sigma^2_{_{{\cal C}_{W_1}}}(x_1) \cdot \sigma^2_{_{{\cal C}_{W_0}}}(x_0)+\dots\; ,
    \quad\textrm{where}\quad
    \sigma^2_{_{{\cal C}_{W_i}}}(x_i) = \Big\langle \Big[W_i(x_i) - \langle
        W_i(x_i)
    \rangle\Big]^2 \Big\rangle,
\end{align}
and the ellipsis stands for exponentially suppressed terms.

A simple example is the disconnected contribution to a two-point function of
fermion bilinears, whose disconnected components are the
single-propagator traces
\begin{align}
    \bar t_{_{\Gamma,r}}(x_0) &= - \frac{a_{_\Gamma}}{aL^3} \sum_{\bm x}\tr \left[\Gamma D^{-1}_{m_r}(x,x) \right]\, ,
\end{align}
where $D_{m_r}$ is the massive Dirac operator with bare-quark mass $m_r$,
$a$ is the lattice spacing and $L^3$ is the lattice volume.
The factor $a_\Gamma=-\i$ for $\Gamma=\gamma_\mu$ and $a_\Gamma=1$ for
$\Gamma=\id,\gamma_5,\gamma_\mu\gamma_5,\sigma_{\mu\nu}$, is chosen so that
the Wick contraction is real.
The gauge variance can be defined in terms of local operators
\begin{align}
    \sigma^2_{\bar t{_{\Gamma,r}}} =  \frac{a^2_{_\Gamma}}{L^3} \sum_{\bf x} a^3 \langle
    O_{_{\Gamma, rr}}(0,{\bf x})\, O_{_{\Gamma, r'r'}}(0)\rangle_c\, ,
\end{align}
where $c$ stands for connected correlation function, $O_{_{\Gamma, rs}}(x) =
\bar\psi_r(x) \Gamma \psi_s(x)$, and $m_{r'}=m_r$.
It is evident that the gauge variance is itself a disconnected contraction,
and so begins only at order $g_0^4$ or higher in perturbation theory and may
be expected to be suppressed.
Nevertheless, by the operator product expansion and power-counting the
variance has a cubic divergence in the continuum limit.

In practice, it is unfeasible to compute the single-propagator trace exactly,
to which end we introduce $N_s$ independent auxiliary fields $\eta_i(x)$,
whose components must have unit variance and zero mean, which we choose to be
drawn from a Gaussian distribution, and a stochastic estimator 
\begin{align}
    \bar\tau_{_{\Gamma,r}}(x) = -\frac{1}{aL^3 N_s}
    \sum_{i=1}^{N_s}\sum_{\bm x}
    {\rm Re} \left[a_{_\Gamma} \eta^\dagger_i(x) \Gamma
    \{D^{-1}_{m_r}\eta_i\}(x)\right]\, .
\end{align}
The variance of the stochastic estimator receives contributions from the
fluctuations of the auxiliary fields
\begin{align}
    \sigma^2_{\bar\tau_{_{\Gamma,r}}} \hspace{-0.25cm} =  
    \sigma^2_{\bar t_{_{\Gamma,r}}} - \frac{1}{2 L^3 N_s}\left\{
        a_{_\Gamma}^2 \sum_{\bf x} a^3 \langle O_{_{\Gamma, rr'}} (0,{\bf x}) O_{_{\Gamma, r'r}} (0) \rangle 
        \hspace{-0.05cm} + \hspace{-0.05cm}\frac{1}{a}
        \sum_x a^4 \langle P_{rr'} (x) P_{r'r} (0) \rangle\right\}\label{eq:tauV2},
\end{align}
where $P_{rs}=O_{_{\gamma_5,rs}}$.
The second and third terms in eq.~\eqref{eq:tauV2} are connected diagrams which
occur at tree level in perturbation theory, which suggests that they will
be larger than the gauge variance unless $N_s$ is large.
Note that the third term, which is chirally-enhanced, is independent of
$\Gamma$.

\begin{table}[t]
\small
\begin{center}
\setlength{\tabcolsep}{.10pc}
\begin{tabular}{@{\extracolsep{0.4cm}}ccccccccc}
\hline
id &$L/a$&$\kappa$&MDU&$N_{\rm cfg}$&$M_\pi$[MeV]&$M_\pi L$\\
\hline
F7  &$48$&$0.13638$ &$9600$&  $100(1200)$ &$268$ &$4.3$ \\
G8  &$64$&$0.136417$&$820$&   $25$ &$193$ &$4.1$ \\
\hline
\end{tabular}
\end{center}
\caption{\label{tab:ens} Overview of the ensembles and statistics presented in this
study and their simulation and physics parameters.
}
\end{table}

\begin{figure}[t]
\begin{center}
\includegraphics[width=0.6\columnwidth]{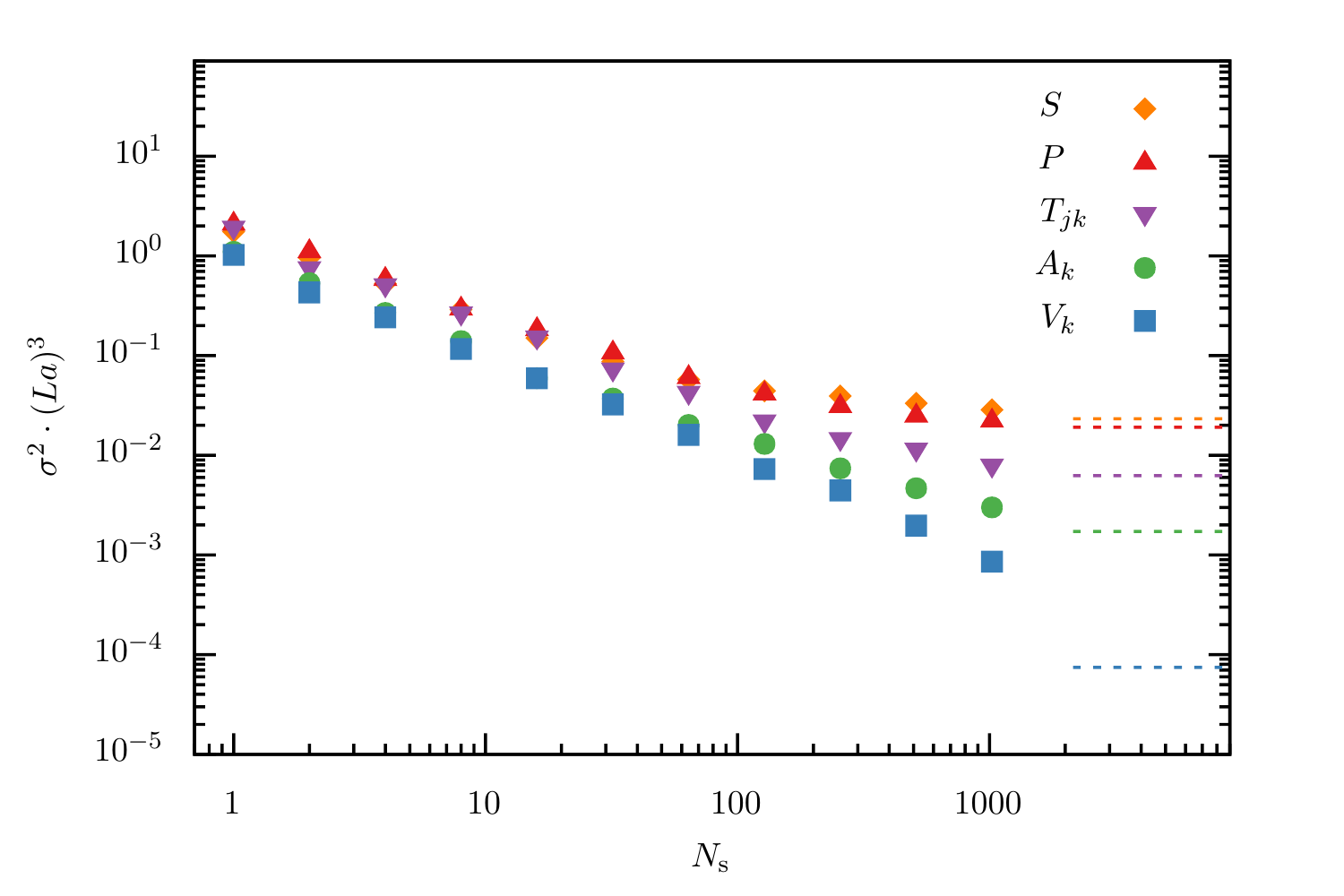}
\caption{Variances of the standard random noise estimators defined in
    eq.~\eqref{eq:tauV2} as a function of $N_s$ for the ensemble F7. The
    symbols $S$, $P$, $T_{jk}$, $A_k$ and $V_k$ stand for $\Gamma=I$,
    $\gamma_5$, $\sigma_{jk}$, $\gamma_k\gamma_5$ and $\gamma_k$ respectively.
    The dashed lines indicate the gauge noise contributions to the variances
    computed in sec.~\ref{sec:freq}. \label{fig:std1}}
\label{fig:onept}
\end{center}
\end{figure}

In order to investigate the relative magnitude of the contributions to the
variance, we employed the ensembles listed in table~\ref{tab:ens}, with
$N_\mathrm{f}=2$ flavours of non-perturbatively O($a$)-improved Wilson
fermions.
In fig.~\ref{fig:onept} we plot the variance as a function of $N_s$ for the
scalar and pseudoscalar densities, and the tensor, axial and vector currents
for the F7 ensemble.
The linear behaviour in $N_s^{-1}$ illustrates that the random noise variance
dominates by orders of magnitude for small $N_s$, and is practically
independent of the bilinear, as expected from the arguments outlined above.
For large $N_s$, the gauge variance (dashed lines) is saturated, and is orders
of magnitude larger for the densities than the currents.
It is therefore highly desirable to examine variance reduction methods in
particular for the currents in order to reach the gauge noise.

\section{Estimators for differences of single-propagator traces}
\label{sec:diff}
In this section, we investigate the difference of single-propagator traces
$\bar t_{_{\Gamma,rs}}=\bar t_{_{\Gamma,r}}-\bar t_{_{\Gamma,s}}$ with two
different bare quark masses, $m_r<m_s$, which contains the infrared
contributions to the single-propagator trace.
Such differences arise, for example, from the disconnected contraction of the
electromagnetic current in the isosymmetric theory, as well as constituting a
component of our improved frequency-splitting estimator in
sec.~\ref{sec:freq}.

We examine two estimators for the differences of single-propagator traces
defined by\footnote{Here, we have used the identity
    $D_{m_r}^{-1}-D_{m_s}^{-1}=(m_s-m_r)D_{m_r}^{-1}D_{m_s}^{-1}$ to rewrite
the difference as a product.}
\begin{align}
    \bar\theta_{_{\Gamma,rs}}(x_0)
    &=  -\frac{(m_s-m_r)}{aL^3 N_s} \sum_{i=1}^{N_s} \sum_{\bm x}{\rm Re} \left[ a_{_\Gamma}\eta^\dagger_i(x) \Gamma
        \{D^{-1}_{m_r}D^{-1}_{m_s}\eta_i\}(x)\right]\, ,\label{eq:theta12}\\
    \bar\tau_{_{\Gamma,rs}}(x_0) &=  -\frac{(m_s-m_r)}{aL^3 N_s} \sum_{i=1}^{N_s}\sum_{\bm x}
        {\rm Re}\left[a_{_\Gamma} \{\eta^\dagger_i  D^{-1}_{m_r}\}(x)\, \Gamma\, \{D^{-1}_{m_s}\eta_i\}(x)\right] \; ,
\end{align}
which we denote as the standard~\cite{Francis:2014hoa} and split-even
estimators~\cite{Giusti:2019kff} in the following.
Note that the cost of the two estimators is the same for a given $N_s$.
Analogously to the previous section, their variances can be defined in terms
of local operators
\begin{align}
    \label{eq:RNV6}
    \sigma^2_{\bar\theta_{_{\Gamma,rs}}} & = \sigma^2_{\bar t_{_{\Gamma,rs}}}-
        \frac{(m_s-m_r)^2}{2 L^3 N_s} \left\{
        a_{_\Gamma}^2\hspace{-0.25cm}\sum_{y_1,{\bf y_2},y_3}\hspace{-0.25cm} a^{11} \langle S_{rs}(y_1) O_{_{\Gamma, ss'}} (0,{\bf y_2}) S_{s'r'}(y_3) O_{_{\Gamma, r'r}} (0) \rangle + 
        \right.\nonumber\\
    & \left.\hspace{3.375cm}
        \frac{1}{a}
        \sum_{y_1,y_2,y_3} a^{12} \langle S_{rs}(y_1) P_{ss'} (y_2) S_{s'r'}(y_3)
        P_{r'r} (0) \rangle\right\}\, ,\\
    \sigma^2_{\bar\tau_{_{\Gamma,rs}}}  & = 
        \sigma^2_{\bar t_{_{\Gamma,rs}}} - \frac{a_{_\Gamma}^2 (m_s-m_r)^2}{2 L^3 N_s}
        \sum_{y_1,{\bf y_2},y_3} a^{11} \Big\{
            \big\langle  S_{rs}(y_1)\, O_{_{\Gamma, ss'}}(0,{\bf y_2})\,
            S_{s'r'}(y_3) \, O_{_{\Gamma, r'r}} (0) \big\rangle \nonumber\\[0.25cm]
    & \hspace{4.0cm} + \big\langle  P_{rr'}(y_1)\, O_{_{\Gamma, r's'}}(0,{\bf y_2})\,
        P_{s's}(y_3) \, O_{_{\Gamma, sr}} (0) \big\rangle 
       \Big\}
    \label{eq:RNV7}
\end{align}
where $S_{rs}=O_{_{\id,rs}}$ and the gauge variance $\sigma^2_{\bar
t_{_{\Gamma,rs}}}$ is the fully-connected analogue of the first four-point
function in eq.~\eqref{eq:RNV6}, just as for the single-propagator trace.
The two estimators differ only between the second four-point functions in
eq.~\eqref{eq:RNV6} and eq.~\eqref{eq:RNV7}.
For the standard estimator, the $\langle{SPSP}\rangle$ term is independent of
$\Gamma$ and integrated over one more time coordinate compared to the
$\langle{POPO}\rangle$ term in the split-even estimator, which furthermore
depends on $\Gamma$.

\begin{figure}[t]
\begin{center}
\includegraphics[width=0.45\columnwidth]{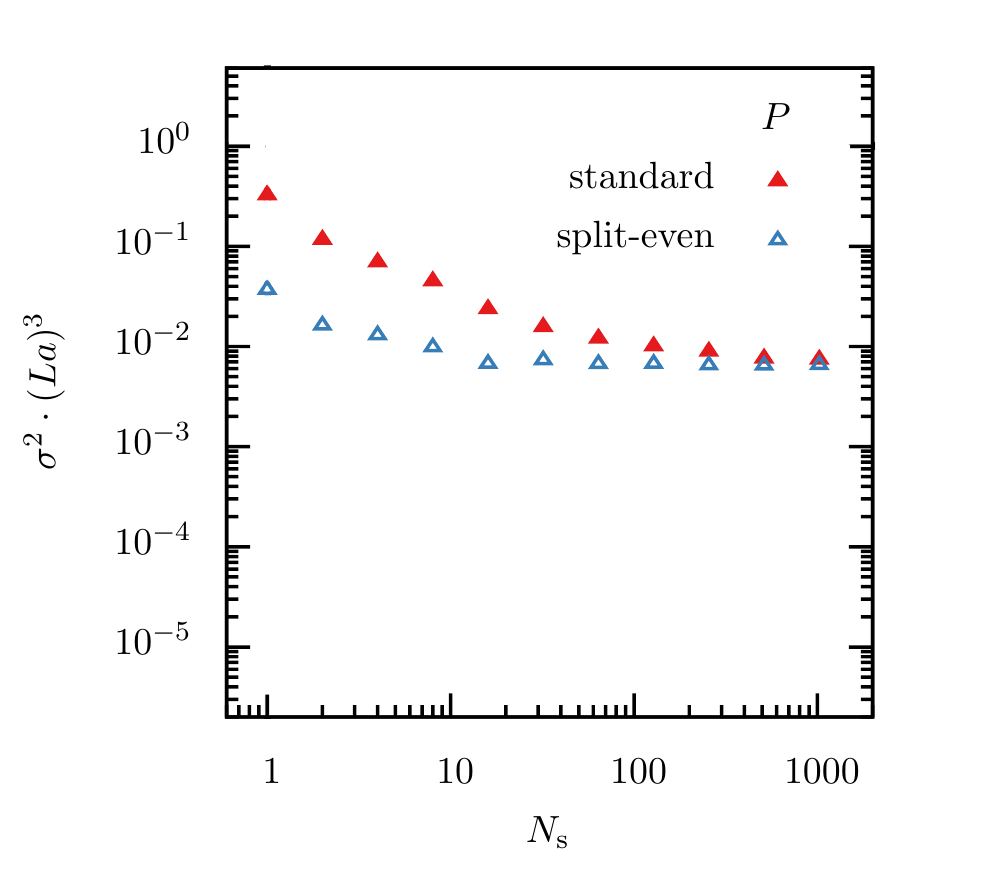}
\includegraphics[width=0.45\columnwidth]{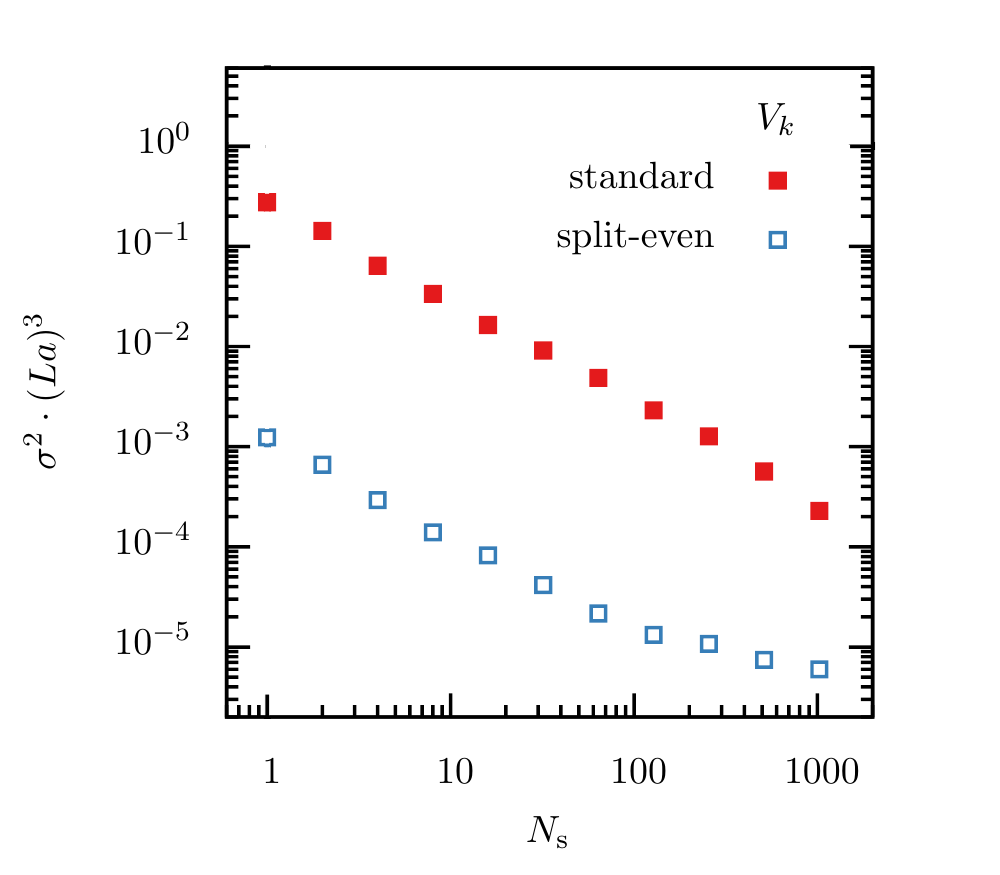}
\caption{Variances of the standard $\theta_{_{\Gamma,rs}}$ (filled symbols) and the
  split-even $\tau_{_{\Gamma,rs}}$ (open symbols) estimators 
  for the pseudoscalar density (left) and the vector current
  (right) for ensemble F7.
  \label{fig:spln1}}
\end{center}
\end{figure}

In fig.~\ref{fig:spln1}, we compare the variances of the standard (filled) and
split-even (open) estimators for the pseudoscalar and vector channels for the
difference of light- and strange-quark propagators with bare-quark masses
$am_{q,r}=0.00207$ and $am_{q,s}=0.0189$ on the F7 ensemble.
The variance of the split-even estimator is one to two orders of magnitude
smaller than the standard one, and reaches the gauge noise with
$N_s\sim\mathrm{O}(10)$ for the pseudoscalar channel, and O(100) in the vector
channel.

Interestingly, the large gain using the split-even estimator reported here
has been confirmed in ref.~\cite{Wittig:2019qbu}.
The partial cancellation of stochastic noise between the light- and
strange-quark traces is already present in the baseline standard estimator,
which is not the origin of the significant gain as suggested there.
The large reduction in the variance is well explained instead by the preceding
formul\ae\,for the variances of the two estimators.
This analysis also explains the origin of the empirical gains observed for the
one-end trick for the pseudoscalar density in twisted-mass
QCD~\cite{Boucaud:2008xu}, in which case it is even possible to show the
estimator has a strictly smaller variance than the standard
one~\cite{Giusti:2019kff}.

\begin{figure}[t]
\begin{center}
\includegraphics[width=0.45\columnwidth]{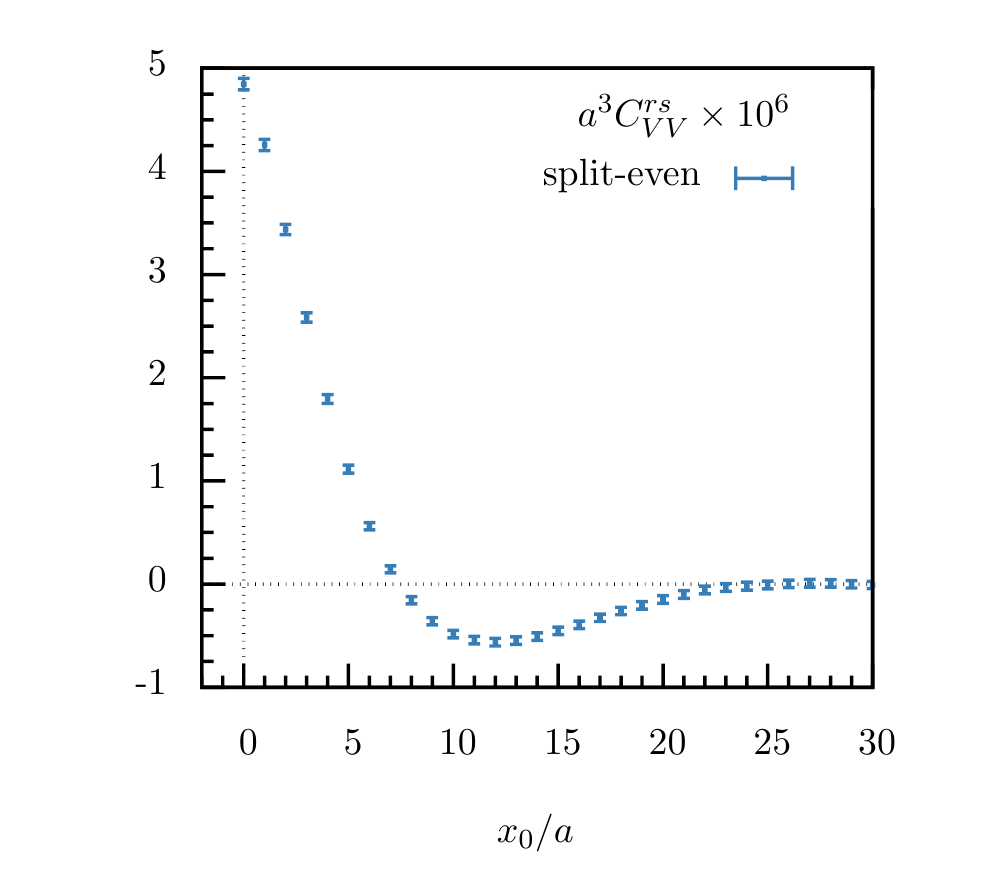}
\includegraphics[width=0.45\columnwidth]{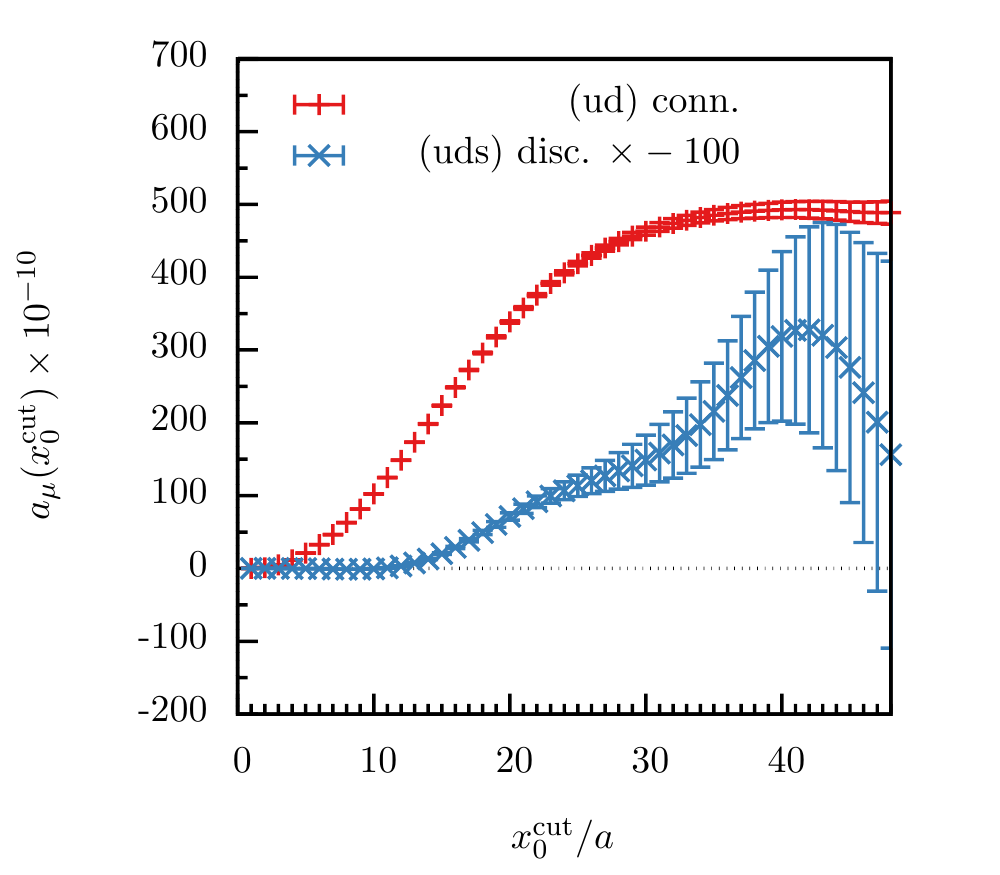}
\caption{Left: the disconnected contribution to the electromagnetic current
    correlator using the split-even estimator from $N_\mathrm{cfg}=1200$ gauge
    configurations with $N_s=512$ noise sources for F7, and (right) the
    corresponding contribution to $a_\mu$ from distances up to
    $x_0^\mathrm{cut}$.  A striking plateau in $x_0^\mathrm{cut}$ is
    visible in the connected case, but no evident for the disconnected
    contribution and moreover, the error grows quickly due to the vacuum
    contributions to the gauge noise, which can be tackled with multi-level
    integration.
}
\label{fig:twopt_em} 
\end{center}
\end{figure}

As an application, we consider the disconnected contribution to the
electromagnetic current correlator for an isodoublet of light quarks and a
valence strange quark, $C^{rs}_{VV}(x_0)\sim\langle\bar
t_{\gamma_k,rs}(x_0+y_0)\,
\bar t_{\gamma_k,rs}(y_0)\rangle$.
The disconnected contribution is shown in fig.~\ref{fig:twopt_em} (left) using
$N_\mathrm{cfg}=1200$, and a good signal is observed up to 1.0~fm.
The current correlator determines the hadronic vacuum polarization
contribution to $a_\mu$, and in fig.~\ref{fig:twopt_em} (right) we show the
resulting disconnected contribution (times a factor 100, for visibility)
computed using the split-even estimator, along with the light-quark connected
contribution, from distances up to $x_0^{\mathrm{cut}}$ using the
time-momentum representation~\cite{Bernecker:2011gh}.
In particular, this variance-reduction technique represents an important
computational advance for precision computations of hadronic contributions to
muon $g-2$, and as suggested in ref.~\cite{Giusti:2019kff}, the use of this
estimator can significantly speed up many other computations such as of
hadronic matrix elements and their strong and electromagnetic isospin-breaking
corrections, see ref.~\cite{Risch:2019xio} for a first application to
isospin-breaking corrections.

\section{Frequency-splitting estimators}
\label{sec:freq}

\begin{figure}[t]
\begin{center}
\includegraphics[width=0.45\columnwidth]{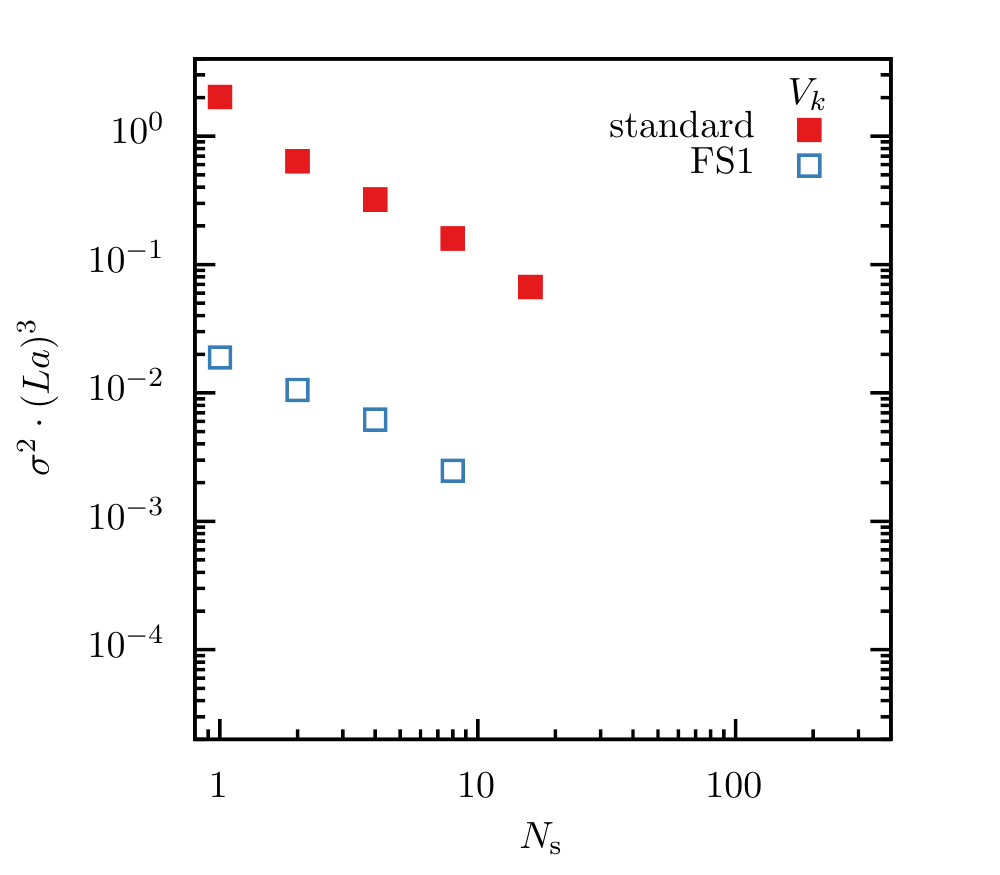}
\includegraphics[width=0.45\columnwidth]{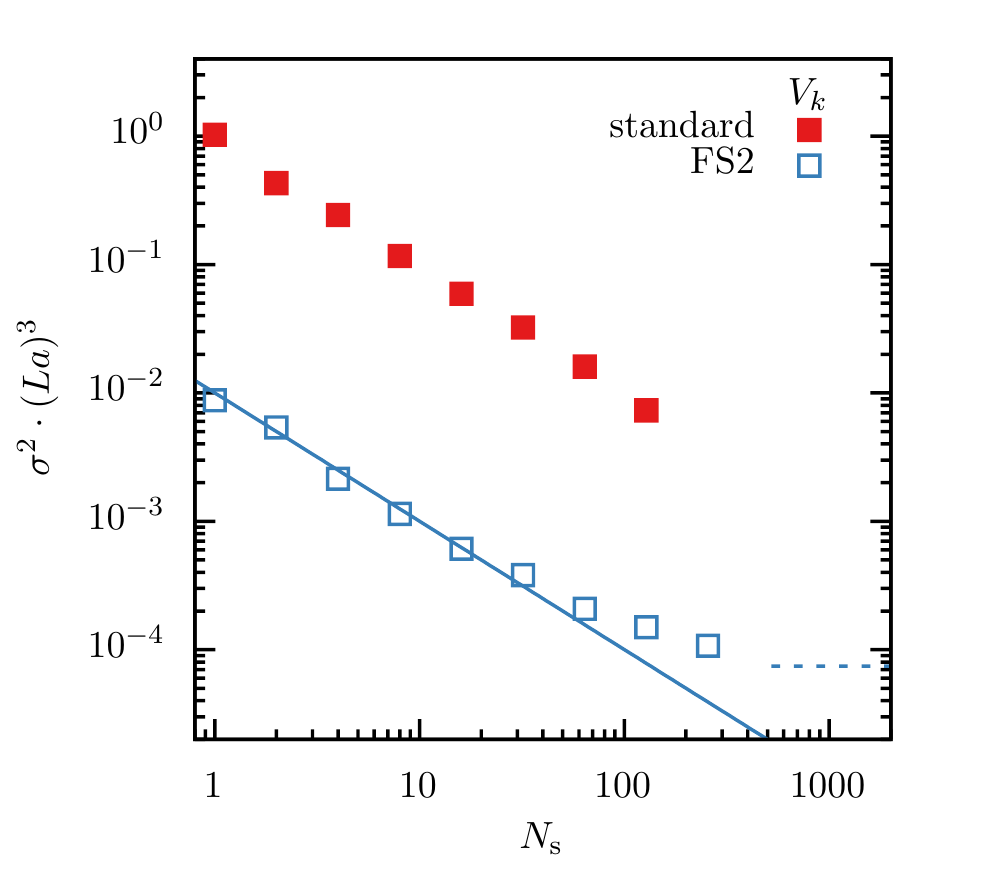}
\caption{Variances of the frequency-splitting estimators (open symbols) for
    the vector current, compared with the standard random-noise estimators
    (filled symbols) on the G8 ensemble (left) and F7 ensemble (right).
    The cost of one iteration of the FS estimator is about 3.3 and 6 times the
    standard one on F7 and G8 respectively.
}
\label{fig:FSEPV}
\end{center}
\end{figure}

In order to construct a reduced-variance estimator for the single-propagator
trace, we combine the split-even estimator and for the large quark-mass
contribution use the order-$2n$ hopping expansion of the Dirac
operator~\cite{Bali:2009hu,Gulpers:2013uca},
$D_{m_r}^{-1} = M_{2n,m_r} + D_{m_r}^{-1} H^{2n}_{m_r}\; ,$   where $H_{m_r}$
denotes the hopping part of the Dirac operator and $M_{2n,{m_r}}$ the first
$2n-1$ terms in the hopping expansion.
We denote the corresponding decomposition of the trace $\bar t_{_{\Gamma,r}} =
\bar t^{M}_{_{\Gamma,r}} + \bar t^{R}_{_{\Gamma,r}}$.
The first term can be computed exactly with $24n^4$ applications of
$M_{2n,m_r}$ onto probing vectors, while an efficient stochastic estimator for
the remainder of the hopping expansion is given by
\begin{align}
    \bar \tau^{R}_{_{\Gamma,r}}(x_0) &= - \frac{1}{a L^3 N_s}
        \sum_{i=1}^{N_s}\sum_{\bf x} {\rm
        Re}\left\{a_{_\Gamma}\big[\eta_{i}^{\dagger} H^{n}_{m_r}\big](x)\,
        \Gamma\, \big[D_{m_r}^{-1}  H^{n}_{m_r} \eta_i\big](x) \right\} \; .
\end{align}
We therefore define the frequency-splitting estimator for the target quark
mass $m_1$ by using the split-even estimator for $K-1$ differences with
$m_{r_k}<m_{r_{k+1}}$ and applying the hopping decompositon at the largest
quark mass $m_{r_K}$ which controls the ultraviolet fluctuations,
\begin{align}
    \bar \tau^{\rm fs}_{\Gamma,r_1}(x_0) = 
    \sum_{k=1}^{K-1} \bar \tau_{_{\Gamma,r_k r_{k+1}}}(x_0) + 
    \bar t^{M}_{\Gamma,r_K}(x_0) + \bar \tau^{R}_{\Gamma,r_K}(x_0)\, ,
\end{align}
In fig.~\ref{fig:FSEPV}, we investigate two frequency-splitting estimators for
single-propagator traces at the sea-quark mass for the G8 (left) and F7
(right) ensemble respectively.
For G8 we use $K=3$ and apply the hopping expansion at $am_{r_K}=0.1$, while
for F7 we take $K=5$ and $am_{r_K}=0.3$, and $n=2$ in both cases%
\footnote{See ref.~\cite{Giusti:2019kff} for the relative $N_s$ used in each
component.}.
In both cases, we see around two orders of magnitude reduction in the
random-noise contributions to the variance, but due to the estimated increase
of about $3.3$ and $6$ in the cost for G8 and F7, the cost reduction in the
vector channel is in the region of $10-30$ depending on the mass.

\section{Conclusions}
\label{sec:conclusions}
In these proceedings, we have investigated improved split-even estimators for
differences of single-propagator traces, which can be applied to efficiently
evaluate the disconnected contributions arising from the electromagnetic
current, so that the leading contribution to the variances arises from
fluctuations of the gauge field.
The cost is reduced by between one to two orders of magnitude depending on
the bilinear.
Furthermore, these can be used to construct frequency-splitting estimators for
single-propagator traces by combining them with the hopping expansion for
large quark masses, which reduces the cost by more than an order of magnitude
for the vector current close to the physical point.
These techniques are compatible with other variance-reduction techniques, such
as low-mode averaging~\cite{Giusti:2004yp,DeGrand:2004qw} and
dilution~\cite{Foley:2005ac}.

{\small
\paragraph{Acknowledgments}
Simulations have been performed on the PC clusters Marconi at CINECA
(CINECA-INFN and CINECA-Bicocca agreements) and Wilson at Milano-Bicocca.
We are grateful to our colleagues within the CLS initiative for
sharing the ensembles of gauge configurations with two dynamical flavours.
L.G. and T. H. acknowledge partial support by the INFN project
``High performance data network''.
}

{\tiny
\printbibliography
}
\end{document}